\documentclass[conference,letterpaper]{IEEEtran}

\usepackage[utf8]{inputenc} 
\usepackage[T1]{fontenc}
\usepackage{url}
\usepackage{ifthen}
\usepackage{cite}
\usepackage[cmex10]{amsmath} % Use the [cmex10] option to ensure complicance
\usepackage{amsmath,amssymb,amsfonts}
\usepackage{algorithmic}
\usepackage{graphicx}
\usepackage{textcomp}
\usepackage{xcolor}
\usepackage{amssymb}
\usepackage{amsmath}
\usepackage{graphicx}
\usepackage{epstopdf}
\usepackage{balance}
\usepackage{epsfig}
\usepackage{graphics}
\usepackage{subfigure}
\usepackage{lineno}
\usepackage{epsfig}
\usepackage{color}
\usepackage{textcomp}
\usepackage{algorithm}  
\usepackage{bm}
\usepackage{upgreek}
\usepackage{filecontents}
\graphicspath{ {Figures/}}
\usepackage{setspace}
\usepackage{hyperref}
\usepackage{acronym}
\acrodef{cas}[CAS]{communication-assisted sensing}
\acrodef{snc}[S\&C]{sensing and communications}
\acrodef{rd}[R-D]{rate-distortion}
\acrodef{iid}[i.i.d.]{independently and identically distributed}
\acrodef{mse}[MSE]{mean squared error}
\acrodef{mmse}[MMSE]{nimimum mean squared error}
\acrodef{glm}[GLM]{Gaussian linear model}
\acrodef{mi}[MI]{mutual information}
\acrodef{ss}[S-side]{Sensing side}
\acrodef{cs}[C-side]{Communication side}
\acrodef{wrt}[w.r.t.]{with respect to}
\acrodef{sct}[SCT]{source-channel separation theorem}
\acrodef{qos}[QoS]{quality of service}
\acrodef{snr}[SNR]{signal to noise ratio}
\acrodef{sw}[SW]{separated S\&C waveforms}
\acrodef{dw}[DW]{dual-functional waveform}

\begin{document}
\title{Fundamental Limits of Communication-Assisted Sensing in ISAC Systems } 

% %%% Single author, or several authors with same affiliation:
% \author{%
%  \IEEEauthorblockN{Andrew R.~Barron}
%  \IEEEauthorblockA{Department of Statistics and Data Science\\
%                    Yale University\\
%                    New Haven, CT, USA\\
%                    Email: andrew.barron@yale.edu}
% }

%%% Several authors with up to three affiliations:
\author{%
  \IEEEauthorblockN{Fuwang Dong$^1$, Fan Liu$^1$, Shihang Lu$^1$, Yifeng Xiong$^2$, Weijie Yuan$^1$, Yuanhao Cui$^1$}
  \IEEEauthorblockA{ 1. Sourthen University of Science and Technology, China\\  
  	                 2. Beijing University of Posts and Telecommunications, China \\              
                        Email: liuf6@sustech.edu.cn}
%  \and
%  \IEEEauthorblockN{Claude E.~Shannon and David Slepian}
%  \IEEEauthorblockA{Bell Telephone Laboratories, Inc.\\ 
%                    Murray Hill, NJ, USA\\
%                    Email: \{csh, dsl\}@bell-labs.com}
}

%%% Many authors with many affiliations:
% \author{%
%   \IEEEauthorblockN{Andrew R.~Barron\IEEEauthorrefmark{1},
%                     Claude E.~Shannon\IEEEauthorrefmark{2},
%                     David Slepian\IEEEauthorrefmark{2},
%                     and Jacob Ziv\IEEEauthorrefmark{2}\IEEEauthorrefmark{3}}
%   \IEEEauthorblockA{\IEEEauthorrefmark{1}%
%                    Department of Statistics and Data Science, Yale University, New Haven, CT, USA,
%                     andrew.barron@yale.edu}
%   \IEEEauthorblockA{\IEEEauthorrefmark{2}%
%                     Bell Telephone Laboratories, Inc.,
%                     Murray Hill, NJ, USA,
%                     \{csh,dsl,jz\}@bell-labs.com}
%   \IEEEauthorblockA{\IEEEauthorrefmark{3}%
%                     Department of Electrical Engineering, Technion---Institute of Technology, Haifa, Israel,
%                     jz@ee.technion.ac.il}
% }

\maketitle

\begin{abstract}
In this paper, we introduce a novel communication-assisted sensing (CAS) framework that explores the potential coordination gains offered by the integrated sensing and communication technique. The CAS system endows users with beyond-line-of-the-sight sensing capabilities, supported by a dual-functional base station that enables simultaneous sensing and communication. To delve into the system's fundamental limits, we characterize the information-theoretic framework of the CAS system in terms of rate-distortion theory. We reveal the achievable overall distortion between the target's state and the reconstructions at the end-user, referred to as the sensing quality of service, within a special case where the distortion metric is separable for sensing and communication processes. As a case study, we employ a typical application to demonstrate distortion minimization under the ISAC signaling strategy, showcasing the potential of CAS in enhancing sensing capabilities.      
\end{abstract}


\section{Introduction}\label{intro}
\subsection{Background and Motivations}
Integrated sensing and communication (ISAC) system is widely acknowledged for its potential to enhance \ac{snc} performance by sharing the use of hardware, spectrum, and signaling strategies \cite{9606831}. Recently, the ISAC system has been officially approved as one of the six critical usage scenarios of 6G by the International Telecommunications Union (ITU). Over the past few decades, substantial research efforts have been directed toward signal processing and waveform design (cf. \cite{9540344,9737357}, and the reference therein). However, the fundamental limits and the resulting performance tradeoff between \ac{snc} have remained longstanding challenges within the research community \cite{9705498}.   

Several groundbreaking studies have recently delved into exploring the fundamental limits within various ISAC system configurations. The authors of \cite{8437621,9785593} consider a scenario where the transmitter (Tx) communicates with a user through a memoryless state-dependent channel, simultaneously estimating the state from generalized feedback. The capacity-distortion-cost tradeoff of this channel is characterized to illustrate the optimal achievable rate for reliable communication while adhering to a preset state estimation distortion. Concurrently, another study in \cite{10147248,10206462} focuses on a general scenario where the Tx senses arbitrary targets rather than the specified communication state, namely, the separated \ac{snc} channels. The deterministic-random tradeoff between \ac{snc} within the ISAC signaling strategy is unveiled by characterizing the Cram\'er-Rao bound (CRB)-communication rate region \cite{10147248}. Subsequently, the work in \cite{10206462} extended this tradeoff to any well-defined sensing distortion metrics.  

% The ISAC signal is employed for both \ac{snc} functions, modeled as a random signal conveying information, which is perfectly known to the sensing receiver (Rx), but unknown to the communication user. 

\begin{figure}[!t]
	\centering
	\includegraphics[width=3.5in]{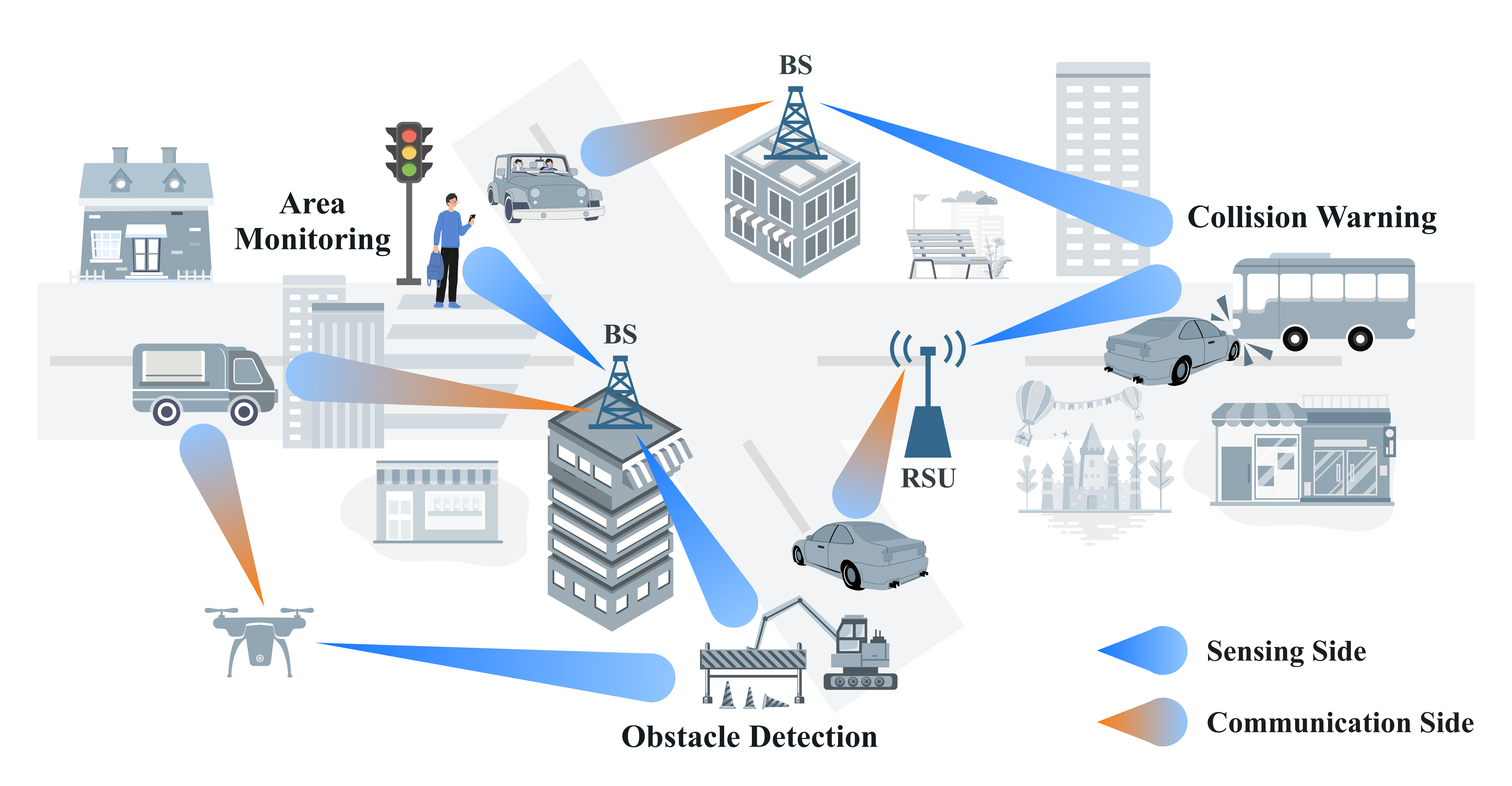}
	\caption{The applications of the proposed novel CAS system.}
	\label{CAS_illustration}
\end{figure}

Unfortunately, there is limited research on exploring the coordination gains offered by mutual support of \ac{snc} functionalities in ISAC systems. Notably, within intelligent 6G applications requiring high-quality sensing service, a natural question arises regarding how sensing performance can be enhanced by incorporating communication functionality, and what the fundamental limits are for such a system. To fill this research gap, we introduce a novel system setup in this paper, which is referred to as \ac{cas}. As illustrated in Fig. \ref{CAS_illustration}, such the \ac{cas} system can endow users with beyond-line-of-sight (BLoS) sensing capability without the need of additional sensors. Specifically, the base station (BS) with favorable visibility illuminates the targets and captures observations containing the interested parameters through device-free wireless sensing abilities, then conveys the relevant information to the users. Thus, the users can acquire the parameters of interest to the BLoS targets.     

The main contributions of this paper are summarized below. First, we establish a novel \ac{cas} system framework, delving deeper into the coordination gain facilitated by the ISAC technique. Second, we analyze the information-theoretic aspects of the \ac{cas} framework, characterizing the fundamental limits of the achievable sensing distortion at the end-user. Finally, we illustrate a practical transmission scheme, designing an ISAC waveform to achieve the minimum sensing distortion.       

\begin{figure*}[!t]
	\centering
	\includegraphics[width=7in]{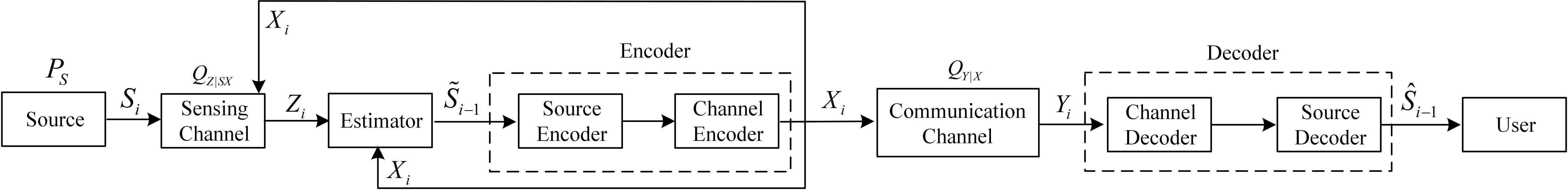}
	\caption{The information-theoretic framework for the CAS systems.}
	\label{fig1}
\end{figure*}

\section{System Model}
\subsection{CAS Process}
The information-theoretic framework of the proposed \ac{cas} system is depicted in Fig. \ref{fig1}. The Tx employs ISAC signal to simultaneously implement \ac{snc} tasks. The state random variables $S$ of the target, the \ac{snc} channel input $X$, the sensing channel output $Z$, and the communication channel output $Y$ take values in the sets $\mathcal{S}$, $\mathcal{X}$, $\mathcal{Z}$, $\mathcal{Y}$, respectively. Here, the state sequence $\{S_i\}_{i \ge 1}$ follows \ac{iid} with a prior distribution $P_S$. The \ac{cas} process mainly contains the following two parts. 

$\bullet$ \textbf{Sensing Process}: The sensing channel output at sensing receiver (Rx) $Z_i$ at a given time $i$ is generated based on the sensing channel law $Q_{Z|SX}(\cdot|x_i,s_i)$ given the $i$th channel input $X_i=x_i$ and the state realization $S_i=s_i$. We assume that the sensing channel output $Z_i$ is independent to the past inputs, outputs and state signals. Let $\tilde{\mathcal{S}}$ denote the estimate alphabet, the state estimator is a map from the acquired the observations $\mathcal{Z}^n$ to $\tilde{\mathcal{S}}^n$. Thus, the expected average per-block estimation distortion in the sensing process can be defined by
\begin{equation}
\Delta_s^{(n)}:= \mathbb{E}[d(S^n,\tilde{S}^n)]=\frac{1}{n}\sum_{i=1}^n \mathbb{E}[d(S_i,\tilde{S}_i)],
\end{equation}    
where $d(\cdot, \cdot)$ is a distortion function bounded by $d_\text{max}$.  

$\bullet$ \textbf{Communication Process}: The Tx encodes the estimate $\tilde{S}_{i-1}$ at the last epoch to communication channel input $X_i$. The communication Rx receives the channel output $Y_i$ according to channel law $Q_{Y|X}$. The decoder is a map from $\mathcal{Y}^n$ to $\hat{\mathcal{S}}^n$, where $\hat{\mathcal{S}}$ denotes the reconstruction alphabet. Thus, the expected average per-block estimation distortion in the communication process can be defined by  
\begin{equation}
\Delta_c^{(n)}:= \mathbb{E}[d(\tilde{S}^n,\hat{S}^n)]=\frac{1}{n}\sum_{i=1}^n \mathbb{E}[d(\tilde{S}_i,\hat{S}_i)].
\end{equation} 

The overall sensing quality of service (QoS) may be measured by the distortion between the source and its reconstruction at the user-end, i.e.,
\begin{equation}
	\Delta^{(n)}:= \mathbb{E}[d(S^n,\hat{S}^n)]=\frac{1}{n}\sum_{i=1}^n \mathbb{E}[d(S_i,\hat{S}_i)].
\end{equation}  

In the \ac{cas} process, a $(2^{n\mathsf{R}},n)$ coding scheme consists of 

1) a state parameter estimator $h$: $\mathcal{X}^n \times \mathcal{Z}^n \to \tilde{\mathcal{S}}^n$;

2) a message set (also the estimate set) $\tilde{\mathcal{S}} = [1:2^{n\mathsf{R}}]$;

3) an encoder $\phi$: $\tilde{\mathcal{S}}^n \to \mathcal{X}^n$;

4) a decoder $\psi$: $\mathcal{Y}^n \to \hat{\mathcal{S}}^n$.

For a practical system, the \ac{snc} channel input $X$ may be restricted by the limited system resources. Define the cost-function $b(x):\mathcal{X} \to \mathbb{R}^+$ to characterize the \ac{snc} channel cost, then a rate-distortion-cost tuple $(\mathsf{R},D,B)$ is said achievable if there exists a sequence of $(2^{n\mathsf{R}},n)$ codes that satisfy
\begin{equation}
	\varlimsup_{n \to \infty} \Delta^{(n)} \le D, \kern 5pt \varlimsup_{n \to \infty} \mathbb{E}[b(X^n)] \le B.
\end{equation}

\textbf{Remark:} We emphasize that while our \ac{cas} system setup shares some similarities with conventional \emph{remote estimation} \cite{kostina2018rate,gao2018optimal,7935515} and \emph{remote source coding} \cite{623151,6915877,8636539} problems, they are essentially distinct. In \emph{remote estimation}, a sensor measures the target's state and transmits its observations to a remote estimator via a wireless channel. In \emph{remote source coding}, the encoder cannot access the original source information but only its noisy observations. Note that in these schemes, the state observations (which may be acquired by additional sensors) are independent of the communication channel input. In contrast, the \ac{snc} channels are coupling due to the employment of the ISAC signaling strategy. For instance, $S \leftrightarrow Z \leftrightarrow X \leftrightarrow Y \leftrightarrow \hat{S}$ forms a Markov chain in the \emph{remote source coding}, which is, however, no longer a Markov chain due to the dependency of observation $Z$ on the dual-functional channel input $X$ in the \ac{cas} framework. As a result, the overall distortion may not be well-defined by applying the remote source coding theory.

\subsection{Sensing Estimator and Constrained Capacity}
In this subsection, we analyze the optimal estimator in the sensing process and the communication channel capacity constrained by the sensing distortion and resource limitation. 

\textit{Lemma 1} \cite{9785593}: By recalling that $S \leftrightarrow XZ \leftrightarrow \tilde{S}$ forms a Markov chain, the sensing distortion $\Delta_s^{(n)}$ is minimized by the deterministic estimator
\begin{equation}\label{opest}
	h^\star(x^n,z^n):= (\tilde{s}^\star(x_1,z_1),\tilde{s}^\star(x_2,z_2),\cdots,\tilde{s}^\star(x_n,z_n)),
\end{equation}   
where
\begin{equation} \label{escost}
	\tilde{s}^\star(x,z):= \arg \min_{s'\in \mathcal{\tilde{S}}} \sum_{s \in \mathcal{S}}Q_{S|XZ}(s|x,z)d(s,s'),
\end{equation}
is independent to the choice of encoder and decoder.

The detailed proof of \textit{Lemma 1} can be found in \cite{9785593}. By applying the estimator \eqref{escost}, we may define the estimate-cost function as $e(x) = \mathbb{E}[d(S,\tilde{s}^\star(X,Z))|X=x]$. Then, a give estimation distortion $D_s$ satisfies 
\begin{equation}
\varlimsup_{n \to \infty} \frac{1}{n}\sum_{i=1}^n \mathbb{E}[e(X)] \le D_s.
\end{equation}
Within the ISAC signaling strategy, the \ac{snc} channel input $X$ may be constrained by both the estimate-cost and resource-cost functions. Namely, the feasible set of the probability distributions of $X$ can be described by the intersection of the two sets as follows.
\begin{equation}
	\mathcal{P}_B=\left\{P_X|\mathbb{E}[b(X)] \le B \right\}, \mathcal{P}_{D_s}=\left\{P_X|\mathbb{E}[e(X)] \le D_s \right\}.
\end{equation}  

\textbf{Definition 1}: In communication process, the information-theoretic capacity constrained by estimation and resource cost is defined by  
\begin{equation}\label{CapTrade}
	C^\text{IT}(D_s, B) = \mathop{\max}\limits_{P_X \in \mathcal{P}_{D_s} \cap \mathcal{P}_B} I(X;Y),
\end{equation}    
where $I(X;Y)$ denotes the \ac{mi} between the communication channel input and output.

From \eqref{CapTrade}, we can observe the coupling relationship between \ac{snc} processes, namely, the achieved sensing distortion concurrently impacts the communication channel capacity. Moreover, the achieved communication distortion $D_c$ is determined by the channel capacity in terms of the rate-distortion theory in lossy data transmission strategy.

\textbf{Definition 2}: The information-theoretic rate distortion function can be defined by 
\begin{equation}\label{srd}
	R^\text{IT}(D_c) = \mathop{\min}\limits_{P(\hat{s}|\tilde{s}): \mathbb{E}[d(\tilde{S},\hat{S})] \le D_c} I(\tilde{S};\hat{S}).
\end{equation}
According to the optimal sensing estimator in Lemma 1, we have $I(\tilde{S};\hat{S}) = I(X,Z;\hat{S}) = I(X;\hat{S})$ due to the fact that $\hat{S}$ is conditional independent to $Z$ with given $X$.  

%$D_c$ represents the communication distortion satisfying $\varlimsup_{n \to \infty} \Delta_c^{(n)} \le D_c$, 

\section{Main Results}

\textit{\textbf{Theorem 1}}: In the \ac{cas} framework, for a separable distortion metric $d(\cdot, \cdot)$ satisfying $\mathbb{E}[d(S,\hat{S})] = \mathbb{E}[d(S,\tilde{S})]+\mathbb{E}[d(\tilde{S},\hat{S})]$, it is possible to design the coding scheme so that the overall sensing QoS is achieved by the sum of estimation and communication distortions $D=D_s+D_c$, if and only if
\begin{equation}\label{theorem}
R^\text{IT}(D_c) \le C^\text{IT}(D_s, B).
\end{equation}

\subsection{Converse}
We start with a converse to show that any achievable coding scheme must satisfy \eqref{theorem}. Consider any $(2^{n \mathsf{R}},n)$ coding scheme defined by the encoding and decoding functions $\phi$ and $\psi$. Let $\tilde{S}^n = h^\star(X^n,Z^n)$ be the estimate sequence as given in \eqref{opest} and $\hat{S}^n = \psi(\phi(\tilde{S}^n))$ be the reconstruction sequence corresponding to $\tilde{S}^n$. 

By recalling from the proof of the converse in lossy source coding, we have
\begin{equation}\label{r1}
\begin{aligned}
\mathsf{R} &\ge \frac{1}{n}\sum_{i=1}^{n} I(\tilde{S}_i;\hat{S}_i) \mathop{\ge} \limits^{(a)}  \frac{1}{n}\sum_{i=1}^{n} R^\text{IT} (\mathbb{E}[d(\tilde{S}_i,\hat{S}_i)]) \\
& \mathop{\ge} \limits^{(b)}  R^\text{IT} \Big(\frac{1}{n} \sum_{i=1}^{n} \mathbb{E}[d(\tilde{S}_i,\hat{S}_i)]\Big) \mathop{\ge} \limits^{(c)} R^\text{IT}(D_c),
\end{aligned}
\end{equation}
where $(a)$ follows the Definition 2, $(b)$ and $(c)$ are due to the convexity and non-increasing properties of the rate distortion function. 

On the other hand, by recalling from the proof of the converse in channel coding, we have
\begin{equation}\label{r2}
\begin{aligned}
\mathsf{R} &\le \frac{1}{n}\sum_{i=1}^{n} I(X_i;Y_i) \\
& \mathop{\le} \limits^{(d)}  \frac{1}{n}\sum_{i=1}^{n} C^\text{IT} \Big(\sum_{x\in\mathcal{X}}P_{X_i}(x)c(x), \sum_{x\in\mathcal{X}}P_{X_i}(x)b(x)\Big) \\
& \mathop{\le} \limits^{(e)}  C^\text{IT} \Big(\frac{1}{n}\sum_{i=1}^{n}\sum_{x\in\mathcal{X}}P_{X_i}(x)c(x), \frac{1}{n}\sum_{i=1}^{n}\sum_{x\in\mathcal{X}}P_{X_i}(x)b(x)\Big) \\
& \mathop{\le} \limits^{(f)} C^\text{IT}(D_s,B),
\end{aligned}
\end{equation}  
where $(d)$ follows the Definition 1, $(e)$ and $(f)$ are due to the concavity and non-decreasing properties of the capacity constrained by estimation and resource costs \cite{9785593}. Additionally, we have the data processing inequality $I(\tilde{S};\hat{S}) = I(X;\hat{S}) \le I(X;Y)$ benefit from the Markov chain $X \leftrightarrow Y \leftrightarrow \hat{S}$. By combing \eqref{r1}, \eqref{r2}, we complete the proof of the converse.

\subsection{Achievability}
Let $S^n$ be drawn \ac{iid} $\sim P_S$, we will show that there exists a coding scheme for a sufficiently large $n$ and rate $\mathsf{R}$, the distortion $\Delta^{n}$ can be achieved by $D$ if \eqref{theorem} holds. The core idea follows the famous \emph{random coding argument} and \emph{source channel separation theorem with distortion}.  

1) \textit{Codebook Generation}: In source coding with rate distortion code, randomly generate a codebook $\mathcal{C}_s$ consisting of $2^{n\mathsf{R}}$ sequences $\hat{S}^n$ which is drawn \ac{iid} $\sim P_{\hat{S}}$. The probability distribution is calculated by $P_{\hat{S}} = \sum_{\tilde{S} }P_{\tilde{S}}P_{\hat{S}|\tilde{S}}$, where $P_{\hat{S}|\tilde{S}}$ achieves the equality in \eqref{srd}. In channel coding, randomly generate a codebook $\mathcal{C}_c$ consisting of $2^{(n\mathsf{R})}$ sequences $X^n$ which is drawn \ac{iid} $\sim P_X$. The $P_X$ is chosen by satisfying the capacity with estimation- and resource-cost in \eqref{CapTrade}. Index the codeword $\tilde{S}^n$ and $X^n$ by $w \in \{1,2,\cdots, 2^{n\mathsf{R}}\}$. 

2) \textit{Encoding}: Encode the $\tilde{S}^n$ by $w$ such that
\begin{equation}
(\tilde{S}^n, \hat{S}^n(w)) \in \mathcal{T}^{(n)}_{d,\epsilon_s} (P_{\tilde{S},\hat{S}}),
\end{equation}
where $\mathcal{T}^{(n)}_{d,\epsilon_s} (P_{\tilde{S},\hat{S}})$ represents the distortion typical set \cite{thomas2006elements} with joint probability distribution $P_{\tilde{S},\hat{S}} = P_{\tilde{S}}P_{\hat{S}|\tilde{S}}$. To send the message $w$, the encoder transmits $x^n(w)$. 

3) \textit{Decoding}: The decoder observes the communication channel output $Y^n=y^n$ and look for the index $\hat{w}$ such that 
\begin{equation}
(x^n(\hat{w}), y^n) \in \mathcal{T}^{(n)}_{\epsilon_c}(P_{X,Y}),
\end{equation} 
where $\mathcal{T}^{(n)}_{\epsilon_c}(P_{X,Y})$ represents the typical set with joint probability distribution $P_{X,Y}=P_XP_{Y|X}$.  If there exists such $\hat{w}$, it declares $\hat{S}^n = \hat{s}^n(\hat{w})$. Otherwise, it declares an error. 

4) \textit{Estimation}: The encoder observes the channel output $Z^n = z^n$, and computes the estimate sequence with the knowledge of channel input $x^n$ by using the estimator $\tilde{s}^n = h^\star(x^n,z^n)$ given in \eqref{opest}. 

5) \textit{Distortion Analysis}: We start by analyzing the expected communication distortion (averaged over the random codebooks, state and channel noise). In lossy source coding, for a fixed codebook $\mathcal{C}_s$ and choice of $\epsilon_s > 0$, the sequence $\tilde{s}^n \in \tilde{\mathcal{S}}^n$ can be divided into two categories:

$\bullet$ $(\tilde{s}^n, \hat{s}^n(w)) \in \mathcal{T}^{(n)}_{d,\epsilon_s}$, we have $d(\tilde{s}^n, \hat{s}^n(w)) < D_c + \epsilon_s$; 

$\bullet$ $(\tilde{s}^n, \hat{s}^n(w)) \notin \mathcal{T}^{(n)}_{d,\epsilon_s}$, we denote $P_{e_s}$ as the total probability of these sequences. Thus, these sequence contribute at most $P_{e_s}d_{\max}$ to the expected distortion since the distortion for any individual sequence is bounded by $d_{\max}$.    

According to the achievability of lossy source coding \cite[Theorem 10.2.1]{thomas2006elements}, we have $P_{e_s}$ tends to zero for sufficiently large $n$ whenever $\mathsf{R} \ge R^\text{IT}(D_c)$. 

In channel coding, the decoder declares an error when the following events occur:

$\bullet$  $(x^n (w), y^n) \notin \mathcal{T}^{(n)}_{\epsilon_c}$; 

$\bullet$  $(x^n (w'), y^n) \in  \mathcal{T}^{(n)}_{\epsilon_c}$, for some $w' \ne w$, we denote $P_{e_c}$ as the probability of the error occurred in decoder. The error decoding contribute at most $P_{e_c}d_{\max}$ to the expected distortion. Similarly, we have $P_{e_c} \to 0$ for $n \to \infty$ whenever $\mathsf{R} \le C^\text{IT}(D_s, B)$ according to channel coding theorem \cite{thomas2006elements}. 

On the other hand, the expected estimation distortion can be upper bounded by
\begin{equation}
\begin{aligned}
&\Delta_s^{(n)} = \frac{1}{n}\sum_{i=1}^n \mathbb{E}[d(S_i,\tilde{S}_i)|\hat{W} \ne w] P(\hat{W} \ne w) \\
& \kern 30pt + \frac{1}{n}\sum_{i=1}^n \mathbb{E}[d(S_i,\tilde{S}_i)|\hat{W} = w] P(\hat{W} = w) \\
& \le P_{e_c}d_\text{max} + \frac{1}{n}\sum_{i=1}^n \mathbb{E}[d(S_i,\tilde{S}_i)|\hat{W} = w] P(\hat{W} = w)(1-P_{e_c}). 
\end{aligned}
\end{equation}

Note that $(s^n, x^n(w), \tilde{s}^n) \in \mathcal{T}^{(n)}_{\epsilon_e} (P_{S,X,\tilde{S}})$ where $P_{S,X,\tilde{S}}$ denotes the joint marginal distribution of $P_{S,X,Z,\tilde{S}} = P_SP_XQ_{Z|SX} \mathbb{I} \{\tilde{s}=\tilde{s}^\star(x,z)\}$ with $\mathbb{I}(\cdot)$ being the indicator function, we have 
\begin{equation}
\varlimsup_{n \to \infty} \frac{1}{n}\sum_{i=1}^n \mathbb{E}[d(S_i,\tilde{S}_i)|\hat{W} = w] \le (1+\epsilon_e) \mathbb{E}[d(S,\tilde{S})],
\end{equation}    
according to the typical average lemma \cite{9785593}. In summary, the overall sensing QoS can be calculated by
\begin{equation}
\begin{aligned}
\Delta^{(n)} & \mathop{=} \limits^{(g)}  \Delta_s^{(n)} + \Delta_c^{(n)} \\
& \mathop{\le} \limits^{(h)} D_c + (P_{e_s}+ 2P_{e_c})d_{\max} + (1+\epsilon_e)(1-P_{e_c})D_s, 
\end{aligned}
\end{equation} 
where $(g)$ follows the definition of the separable distortion metric in Theorem 1 and in $(h)$ we omit the terms containing the product of $P_{e_s}$ and $P_{e_c}$. Consequently, taking $n \to \infty$ and $P_{e_s}, P_{e_c}, \epsilon_c,\epsilon_e,\epsilon_s \to 0$, we can conclude that the expected sensing distortion (averaged over the random codebooks, state and channel noise) tends to be $D \to D_c+D_s$ whenever $ R^\text{IT}(D_c) \le \mathsf{R} \le C^\text{IT}(D_s, B)$.     

\section{Example}
In this section, we take a ISAC waveform design scheme as an example, where the overall sensing QoS $D$ is minimized by appropriately choosing \ac{snc} channel input. The associated optimization problem can be stated by \footnote{Here, we use the symbol $I(D_s,B)$ instead of $C(D_s,B)$ to emphasize that the optimal distribution does not necessarily achieve the channel capacity.}
\begin{equation} \label{p0}
\begin{aligned}
\mathop { \text{min} }\limits_{P_X} \kern 5pt & D = D_c + D_s   \\
\text{subject to} \kern 5pt & R(D_c) \le I(D_s, B). 
\end{aligned} 
\end{equation} 

The widely employed \ac{glm} for \ac{snc} received signal are given by
\begin{equation}\label{glm1}
\mathbf{Z}=\mathbf{H}_s\mathbf{X}+\mathbf{N}_s, \kern 5pt  \mathbf{Y}=\mathbf{H}_c\mathbf{X}+\mathbf{N}_c,
\end{equation}  
where $\mathbf{Z} \in \mathbb{C}^{M_s \times T}$ and $\mathbf{Y} \in \mathbb{C}^{M_c \times T}$ represent the received signals of \ac{snc} Rxs,  $\mathbf{X}\in \mathbb{C}^{N \times T}$ is the transmitting ISAC signal with $T$ being the number of symbols. $N$, $M_s$, and $M_c$ are the numbers of antennas for Tx, sensing Rx, and user Rx, respectively. $\mathbf{H}_s\in \mathbb{C}^{M_s \times N}$ and $\mathbf{H}_c\in \mathbb{C}^{M_c \times N}$ denote the \ac{snc} channels. $\mathbf{N}_s$ and $\mathbf{N}_c$ are the corresponding channel noises whose entries follow the complex Gaussian distribution with $\mathcal{CN}(0,\sigma_s^2)$ and $\mathcal{CN}(0,\sigma_c^2)$.     

Let us consider the scenario of target response matrix (TRM) estimation, where the user wants to acquire the sensing channel information $\mathbf{s} = \text{vec}(\mathbf{H}_s)$ through the \ac{cas} process. Vectorize the Hermitian of sensing received signal by 
\begin{equation}\label{vecs}
\mathbf{z}_s=\tilde{\mathbf{X}}\mathbf{s}+\mathbf{n}_s,
\end{equation}  
where $\mathbf{z}_s=\text{vec}(\mathbf{Z}_s^H)$, $\tilde{\mathbf{X}}=\mathbf{I}_{M_s} \otimes \mathbf{X}^H$, $\mathbf{n}_s = \text{vec}(\mathbf{N}_s^H)$. For discussion convenience, we assume that the parameter vector $\mathbf{s}$ follows Gaussian distribution $\mathcal{CN}(0,\mathbf{I}_{M_s} \otimes \bm{\Sigma}_s)$, where $\bm{\Sigma}_s$ denotes the covariance matrix of each column of $\mathbf{H}_s$ \cite{8579200}.   

\subsection{Sensing Process}
By applying the estimator \eqref{escost}, the \ac{mse} between $\mathbf{s}$ and its estimate $\tilde{\mathbf{s}}$ may be obtained by \cite{BookEstimationTheory}  
\begin{equation}\label{mmse1}
	D_s = \mathbb{E}\Big[\|\mathbf{s}-\tilde{\mathbf{s}}\|^2\Big] = M_s\text{Tr} \Big[ \Big( \frac{T}{\sigma^2_s} \mathbf{X}\mathbf{X}^H + \bm{\Sigma}_s^{-1} \Big)^{-1} \Big].
\end{equation}

Fortunately, when we adopt \ac{mse}, a widely used metric in parameter estimation, as the distortion metric, the condition of separable distortion in the \ac{cas} process holds. Namely, the overall distortion $D$ can be equivalently decomposed into the sum of the estimation and communication distortions,
\begin{equation} {\label{Dsc2}} 
	\begin{aligned}	
		D&=\mathbb{E}\left[\left\| \mathbf{s}-\hat{\mathbf{s}} \right\|_2^2\right] = \mathbb{E}\left[\left\| \mathbf{s}- \tilde{\mathbf{s}} + \tilde{\mathbf{s}} - \hat{\mathbf{s}} \right\|_2^2\right] \\
		&\mathop = \limits^{(i)} \mathbb{E}\left[\left\|\mathbf{s}- \tilde{\mathbf{s}}\right\|_2^2 \right]+\mathbb{E}\left[\left\|\tilde{\mathbf{s}} - \hat{\mathbf{s}}\right\|_2^2 \right] \mathop = \limits^{\Delta} D_s+D_c,
	\end{aligned}
\end{equation}     
where $(i)$ holds from the properties of the conditional expectation
\begin{equation}
\mathbb{E}\big[(\mathbf{s}- \tilde{\mathbf{s}})^T(\tilde{\mathbf{s}} - \hat{\mathbf{s}})\big]= \mathbb{E}\big[(\mathbf{s}-\mathbb{E}\left[ \mathbf{s}|\mathbf{z}, \mathbf{x} \right])^Tf(\mathbf{z}, \mathbf{x})\big]=0. \nonumber
\end{equation}
Here, $\tilde{\mathbf{s}} = \mathbb{E}\left[ \mathbf{s}|\mathbf{z}, \mathbf{x} \right]$ is the \ac{mmse} estimator, and $f(\mathbf{z}, \mathbf{x})$ represents an arbitrary function with respect to $\mathbf{z}, \mathbf{x}$. Furthermore, the \ac{mmse} estimate $\tilde{\mathbf{s}}$ can be expressed by \cite{BookEstimationTheory} 
\begin{equation}
	\tilde{\mathbf{s}} = \Big(\mathbf{I}_{M_s} \otimes \big(\bm{\Sigma}_s\mathbf{X}\mathbf{R}_z^{-1}\big)\Big)\textbf{z},
\end{equation} 
where $\mathbf{I}_{M_s} \otimes \mathbf{R}_z$ is the covariance matrix of the observation $\mathbf{z}$ with the expression of $ \mathbf{R}_z =\mathbf{X}^H \bm{\Sigma}_s \mathbf{X} + \sigma^2_s\mathbf{I}_T$.

\subsection{Communication Process}
The estimate $\tilde{\mathbf{s}}$ also follows complex Gaussian distribution $ \mathcal{CN}(\mathbf{0},\mathbf{R}_{\tilde{\mathbf{s}}})$, with the covariance matrix of 
\begin{equation}\label{reta}
\mathbf{R}_{\tilde{\mathbf{s}}} = \mathbf{I}_{M_s} \otimes \bm{\Sigma}_s\textbf{X}\mathbf{R}_z^{-1}\textbf{X}^H\bm{\Sigma}_s^H.
\end{equation} 
Based on the rate distortion function for the Gaussian distribution source, we have 
\begin{equation} \label{rdcc}
	R(D_c) = \sum_{i=1}^{NM_s} \log \frac{\lambda_i(\mathbf{R}_{\tilde{\mathbf{s}}})}{D_{c_i}}, 
\end{equation}     
which is in a reverse water-filling form as
\begin{equation}
	D_c = \sum_{i=1}^{NM_s}D_{c_i}=\sum_{i=1}^{NM_s} \lambda_i(\mathbf{R}_{\tilde{\mathbf{s}}})-\big(\lambda_i(\mathbf{R}_{\tilde{\mathbf{s}}})-\xi\big)^+,
\end{equation} 
where $\lambda_i(\mathbf{R}_{\tilde{\mathbf{s}}})$ represents the $i$th eigenvalue of $\mathbf{R}_{\tilde{\mathbf{s}}}$, and $\xi$ is the reverse water-filling factor. 

Our aim is to optimize the probability distribution $P_X$. However, there may be no explicit expression of the communication channel capacity for an arbitrary distribution of channel input. To circumvent numerical computing, we temporarily restrict the waveform to the Gaussian distribution with $\mathcal{CN}(\mathbf{0},\mathbf{R}_x)$ which is widely considered in communication systems. It should be noted that although Gaussian distribution is optimal for communication process, but not necessarily achieving optimal estimation performance, leading to an uncertain overall sensing QoS in the \ac{cas} system. Evidently, this also exhibits a deterministic-random tradeoff discussed in \cite{10147248,10206462}. 

The statistical covariance matrix $\mathbf{R}_x$ can be approximated by sample covariance $1/T \mathbf{X}\mathbf{X}^H$. Thus, the \ac{mi} between the received and transmitted signals conditioned on communication channel $\mathbf{H}_c$ can be approximately characterized by
\begin{equation} \label{cap}
	I(\mathbf{Y};\mathbf{X}|\mathbf{H}_c) = 
	\log\Big|\frac{T}{\sigma^2_c}\mathbf{H}_c\mathbf{X}\mathbf{X}^H\mathbf{H}_c^H+\mathbf{I}_N\Big|.
\end{equation}  
The original problem \eqref{p0} can be reformulated by 

\begin{equation} \label{p4}
\begin{aligned}
\mathop { \text{min} }\limits_{\mathbf{X}} \kern 3pt & D_s+ D_c   \\
\text{s.t.} \kern 5pt & R(D_c) \le \log\Big|\frac{T}{\sigma^2_c}\mathbf{H}_c\mathbf{X}\mathbf{X}^H\mathbf{H}_c^H+\mathbf{I}_N\Big|, \\
&	D_s = M_s\text{Tr} \Big[ \Big( \frac{T}{\sigma^2_s} \mathbf{X}\mathbf{X}^H + \bm{\Sigma}_s^{-1} \Big)^{-1} \Big], \\
&\eqref{rdcc}, \kern 5pt \text{Tr}(\mathbf{X}\mathbf{X}^H)\le TP_T,
\end{aligned} 
\end{equation}
where $P_T$ is the power budget. We derive the explicit expressions of \ac{mi} and rate distortion function in the scenario of TRM estimation. However, it is still challenging to solve the non-convex problem \eqref{p4} since the imposed reverse water-filling constraint introduces an unknown nuisance parameter (i.e., the factor $\xi$), which may only be determined through a numerical search algorithm in general. The solution of \eqref{p4} is omitted in this paper due to the limited space. We refer the interested reader to \cite{dong2023communication} for more details.  

\subsection{Numerical Results}
We conduct a comparative analysis of the performance of two signaling strategies: the \ac{sw} scheme and the ISAC scheme. In the \ac{sw} scheme, optimal \ac{snc} waveforms are independently selected for the \ac{snc} processes. The \ac{sw} scheme is more similar to the remote estimation due to the decoupling of the sensing observations from the communication channel input. Unlike remote estimation, which typically employs additional sensors for observation collection, the \ac{sw} scheme detect the target through wireless sensing. Consequently, the \ac{snc} resource competition exists, defined by $\text{Tr}(\mathbf{X}_s\mathbf{X}_s^H)+\text{Tr}(\mathbf{X}_c\mathbf{X}_c^H)\le TP_T$ in this paper, due to the shared use of hardware platforms.

\begin{figure}[!t]
	\centering
	\includegraphics[width=3in]{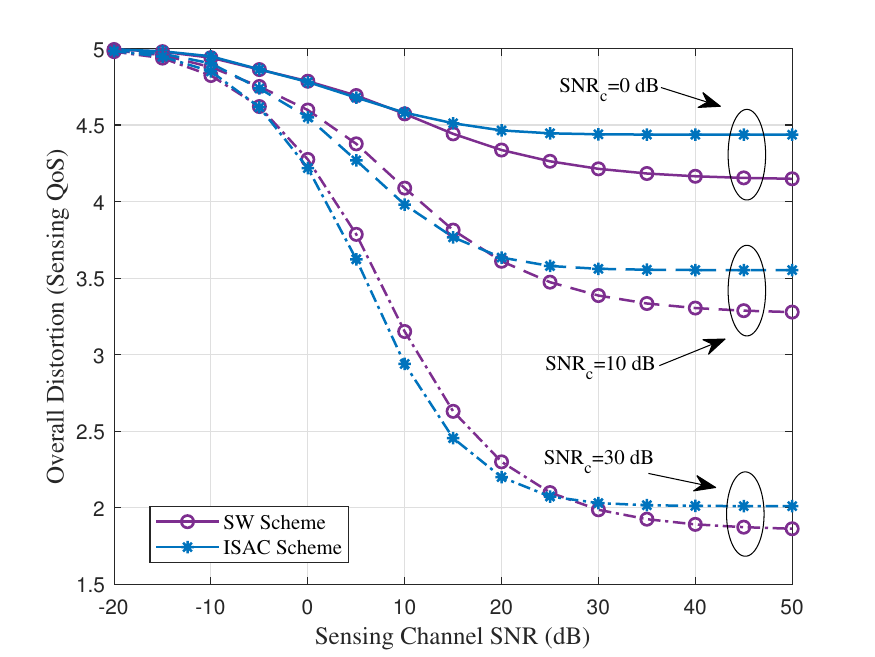}
	\caption{The comparison of the \ac{sw} and ISAC schemes.}
	\label{Compare_SD}
\end{figure}

We can observe from Fig. \ref{Compare_SD} that the power competition between the \ac{snc} processes dominates the overall sensing QoS in the regimes of low SNRs. As anticipated, the ISAC scheme outperforms the \ac{sw} scheme due to the benefits derived from the resource multiplexing gains. Conversely, in high SNR regimes, the \ac{sw} scheme exhibits superior performance compared to the ISAC scheme, implying favorable quality of the \ac{snc} channels. As resource multiplexing gains approach saturation levels, the optimal waveform structures take precedence in influencing the sensing QoS. Specifically, achieving a balance between \ac{snc} channel for the eigenspace of the waveform becomes crucial.

\section{Conclusion}

In this paper, we present a novel communication-assisted sensing (CAS) system framework that endows users with beyond-line-of-sight sensing capabilities. By evaluating the CAS system from an information-theoretic perspective, we characterize the achievable overall distortion between the target's state and its reconstruction at the end-user in a special case where the adopted distortion is separable for sensing and communication processes. We also provide an example of ISAC waveform design, showcasing its superiority over separated sensing and communication waveforms due to power multiplexing gain, especially at low SNR regimes.

%%%%%%
%% Appendix:
%% If needed a single appendix is created by
%%
%\appendix
%%
%% If several appendices are needed, then the command
%%
% \appendices
%%
%% in combination with further \section commands can be used.
%%%%%%
%\IEEEtriggeratref{4}

\bibliographystyle{IEEEtran}
\bibliography{IEEEabrv,CAS_isit24}

\begin{thebibliography}{10}
\providecommand{\url}[1]{#1}
\csname url@samestyle\endcsname
\providecommand{\newblock}{\relax}
\providecommand{\bibinfo}[2]{#2}
\providecommand{\BIBentrySTDinterwordspacing}{\spaceskip=0pt\relax}
\providecommand{\BIBentryALTinterwordstretchfactor}{4}
\providecommand{\BIBentryALTinterwordspacing}{\spaceskip=\fontdimen2\font plus
\BIBentryALTinterwordstretchfactor\fontdimen3\font minus
  \fontdimen4\font\relax}
\providecommand{\BIBforeignlanguage}[2]{{%
\expandafter\ifx\csname l@#1\endcsname\relax
\typeout{** WARNING: IEEEtran.bst: No hyphenation pattern has been}%
\typeout{** loaded for the language `#1'. Using the pattern for}%
\typeout{** the default language instead.}%
\else
\language=\csname l@#1\endcsname
\fi
#2}}
\providecommand{\BIBdecl}{\relax}
\BIBdecl

\bibitem{9606831}
Y.~Cui, F.~Liu, X.~Jing, and J.~Mu, ``Integrating sensing and communications
  for ubiquitous iot: Applications, trends, and challenges,'' \emph{IEEE
  Network}, vol.~35, no.~5, pp. 158--167, 2021.

\bibitem{9540344}
J.~A. Zhang, F.~Liu, C.~Masouros, R.~W. Heath, Z.~Feng, L.~Zheng, and
  A.~Petropulu, ``An overview of signal processing techniques for joint
  communication and radar sensing,'' \emph{IEEE Journal of Selected Topics in
  Signal Processing}, vol.~15, no.~6, pp. 1295--1315, 2021.

\bibitem{9737357}
F.~Liu, Y.~Cui, C.~Masouros, J.~Xu, T.~X. Han, Y.~C. Eldar, and S.~Buzzi,
  ``Integrated sensing and communications: Toward dual-functional wireless
  networks for 6g and beyond,'' \emph{IEEE Journal on Selected Areas in
  Communications}, vol.~40, no.~6, pp. 1728--1767, 2022.

\bibitem{9705498}
A.~Liu, Z.~Huang, M.~Li, Y.~Wan, W.~Li, T.~X. Han, C.~Liu, R.~Du, D.~K.~P. Tan,
  J.~Lu, Y.~Shen, F.~Colone, and K.~Chetty, ``A survey on fundamental limits of
  integrated sensing and communication,'' \emph{IEEE Communications Surveys \&
  Tutorials}, vol.~24, no.~2, pp. 994--1034, 2022.

\bibitem{8437621}
M.~Kobayashi, G.~Caire, and G.~Kramer, ``Joint state sensing and communication:
  Optimal tradeoff for a memoryless case,'' in \emph{2018 IEEE International
  Symposium on Information Theory (ISIT)}, 2018, pp. 111--115.

\bibitem{9785593}
M.~Ahmadipour, M.~Kobayashi, M.~Wigger, and G.~Caire, ``An
  information-theoretic approach to joint sensing and communication,''
  \emph{IEEE Transactions on Information Theory}, pp. 1--1, 2022.

\bibitem{10147248}
Y.~Xiong, F.~Liu, Y.~Cui, W.~Yuan, T.~X. Han, and G.~Caire, ``On the
  fundamental tradeoff of integrated sensing and communications under gaussian
  channels,'' \emph{IEEE Transactions on Information Theory}, vol.~69, no.~9,
  pp. 5723--5751, 2023.

\bibitem{10206462}
F.~Liu, Y.~Xiong, K.~Wan, T.~X. Han, and G.~Caire, ``Deterministic-random
  tradeoff of integrated sensing and communications in gaussian channels: A
  rate-distortion perspective,'' in \emph{2023 IEEE International Symposium on
  Information Theory (ISIT)}, 2023, pp. 2326--2331.

\bibitem{kostina2018rate}
V.~Kostina and B.~Hassibi, ``Rate-cost tradeoffs in scalar lqg control and
  tracking with side information,'' in \emph{2018 56th Annual Allerton
  Conference on Communication, Control, and Computing (Allerton)}.\hskip 1em
  plus 0.5em minus 0.4em\relax IEEE, 2018, pp. 421--428.

\bibitem{gao2018optimal}
X.~Gao, E.~Akyol, and T.~Ba{\c{s}}ar, ``Optimal communication scheduling and
  remote estimation over an additive noise channel,'' \emph{Automatica},
  vol.~88, pp. 57--69, 2018.

\bibitem{7935515}
X.~Ren, J.~Wu, K.~H. Johansson, G.~Shi, and L.~Shi, ``Infinite horizon optimal
  transmission power control for remote state estimation over fading
  channels,'' \emph{IEEE Transactions on Automatic Control}, vol.~63, no.~1,
  pp. 85--100, 2018.

\bibitem{623151}
H.~Viswanathan and T.~Berger, ``The quadratic {G}aussian {CEO} problem,''
  \emph{IEEE Transactions on Information Theory}, vol.~43, no.~5, pp.
  1549--1559, 1997.

\bibitem{6915877}
Y.~Oohama, ``Indirect and direct {G}aussian distributed source coding
  problems,'' \emph{IEEE Transactions on Information Theory}, vol.~60, no.~12,
  pp. 7506--7539, 2014.

\bibitem{8636539}
K.~Eswaran and M.~Gastpar, ``Remote source coding under {G}aussian noise:
  Dueling roles of power and entropy power,'' \emph{IEEE Transactions on
  Information Theory}, vol.~65, no.~7, pp. 4486--4498, 2019.

\bibitem{thomas2006elements}
M.~Thomas and A.~T. Joy, \emph{Elements of information theory}.\hskip 1em plus
  0.5em minus 0.4em\relax Wiley Interscience, 2006.

\bibitem{8579200}
B.~Tang and J.~Li, ``Spectrally constrained {MIMO} radar waveform design based
  on mutual information,'' \emph{IEEE Transactions on Signal Processing},
  vol.~67, no.~3, pp. 821--834, 2019.

\bibitem{BookEstimationTheory}
S.~M. Kay, \emph{Fundamentals of statistical signal processing: Estimation
  theory}.\hskip 1em plus 0.5em minus 0.4em\relax Englewood Cliffs, NJ:
  Prentice-Hall, 1993.

\bibitem{dong2023communication}
F.~Dong, F.~Liu, S.~Lu, Y.~Xiong, Q.~Zhang, Z.~Feng, and F.~Gao,
  ``Communication-assisted sensing in 6{G} networks,'' \emph{arXiv preprint
  arXiv:2311.07157}, 2023.

\end{thebibliography}

\end{document}